\documentclass[twoside,a4paper,11pt]{proceedings}

\usepackage{graphicx}
\usepackage{hyperref}
\usepackage{movie15}
\usepackage{natbib}
\newcommand{\figwidth}{7.6cm}
\begin{document}
\pagenumbering{arabic}
\pagestyle{myheadings}
\thispagestyle{empty}
\vspace*{-1cm}
{\flushleft\includegraphics[width=3cm,viewport=0 -30 200 -20]{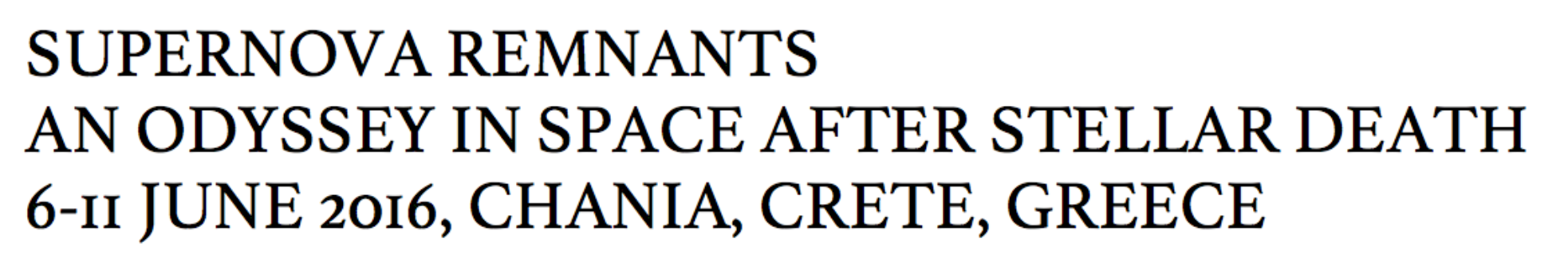}}
\vspace*{0.2cm}
\begin{flushleft}
{\bf {\LARGE
3D simulations of young core-collapse supernova remnants undergoing efficient particle acceleration
}\\
\vspace*{1cm}
Gilles Ferrand$^{1,2}$ and
Samar Safi-Harb$^1$
%
}\\
\vspace*{0.5cm}
%
$^{1}$
Department of Physics and Astronomy, University of Manitoba, Canada \\
$^{2}$
new affiliation: Astrophysical Big Bang Laboratory, Riken, Japan\\
%
\end{flushleft}
\markboth{
3D simulations of CC SNRs with particle acceleration
}{
Ferrand \& Safi-Harb
}
\thispagestyle{empty}
\vspace*{0.4cm}
\begin{minipage}[l]{0.09\textwidth}
\ 
\end{minipage}
\begin{minipage}[r]{0.9\textwidth}
\vspace{1cm}
\section*{Abstract}{\small
Within our Galaxy, supernova remnants are believed to be the major sources of cosmic rays up to the “knee”. However important questions remain regarding the share of the hadronic and leptonic components, and the fraction of the supernova energy channelled into these components. We address such question by the means of numerical simulations that combine a hydrodynamic treatment of the shock wave with a kinetic treatment of particle acceleration. Performing 3D simulations allows us to produce synthetic projected maps and spectra of the thermal and non-thermal emission, that can be compared with multi-wavelength observations (in radio, X-rays, and gamma-rays).
Supernovae come in different types, and although their energy budget is of the same order, their remnants have different properties, and so may contribute in different ways to the pool of Galactic cosmic-rays. Our first simulations were focused on thermonuclear supernovae, like Tycho's SNR, that usually occur in a mostly undisturbed medium. Here we present our 3D simulations of core-collapse supernovae, like the Cas~A SNR, that occur in a more complex medium bearing the imprint of the wind of the progenitor star.
\vspace{10mm}
\normalsize}
\end{minipage}

\section{Introduction}
\label{sec:intro}

Supernova remnants (SNRs) are believed to be the major sources of Galactic cosmic rays (CRs), although their efficiency in accelerating protons -- that make most of the CRs, but are difficult to detect -- is not firmly established. Understanding the origin of Galactic CRs requires a joint modeling of the remnant evolution and of particle acceleration by shock waves. We have developed a framework for simulating this in a 3-dimensional and time-dependent manner, which allows us to produce realistic synthetic maps of the broadband emission from the SNR including the effect of efficient particle acceleration at the blast wave \citep{Ferrand2010a, Ferrand2012g, Ferrand2014b}.
We applied our simulations previously to the case of the remnant from a typical thermo-nuclear (TN) supernova (Type Ia), evolving in a uniform medium. We showed how energetic protons affect the emission from the remnant: they impact the dynamics of the shock wave, and therefore the thermal emission from the shell (in optical and X-rays), and they impact the evolution of the magnetic field, and therefore the non-thermal emission from the electrons (in radio to X-rays and $\gamma$-rays). 
Here we present our first results for the case of the remnant of a core-collapse (CC) supernova (type II or Ib/c), that is still evolving inside the wind of its progenitor. While our previous results were most relevant to an object like Tycho's SNR (G120.1+1.4), our new simulations are more appropriate for an object like Cas~A SNR (G111.7-02.1). 

\section{Model}
\label{sec:model}

We refer the reader to our previous papers for the details. In our approach we combine a hydrodynamic treatment of the blast wave with a kinetic treatment of particle acceleration. The  particles are injected at the shock front from the shocked ambient medium, and then are accelerated by the Fermi~I process. If acceleration is efficient, their pressure can be large enough that it modifies the structure of the shock wave. The diffusive shock acceleration mechanism relies on magnetic turbulence, that scatters particle off -- this effect is embedded inside the diffusion coefficient. Particles that stream in the upstream region can sustain the turbulence they need to keep being accelerated, hence a second back-reaction loop. 

To study this complex system, we rely on numerical simulations. We start our simulations at a small age (here of 5~yr) using self-similar profiles from \cite{Chevalier1983a}. The SNR is then evolved (here up to 300~yr) using a custom version of the code RAMSES \citep{Teyssier2002a}, in a Cartesian grid that is comoving with the blast wave and adaptively refined. Particle acceleration at the forward shock is estimated at each simulation step using the  semi-analytical kinetic model from \cite{Blasi2004a} with extensions from \cite{Caprioli2009a}.

\subsection{The stellar wind}
\label{sec:model_wind}

We consider a fiducial CC SN with explosion energy of $10^{51}~{\rm erg}$ and ejecta mass of $3~M_\odot$, with the ejecta following a steep power-law $\propto r^{-9}$ as a function of radius. For the stellar wind of the progenitor, we adopt a mass loss rate of $\dot{M}_w = 10^{-5}~M_\odot\:\rm{yr^{-1}}$ and a speed of $v_w = 10~\rm{km~s^{-1}}$, parameters that are typical of the red super-giant (RSG) phase and thus appropriate for Cas~A. Conservation of mass implies that the wind profile goes as $r^{-2}$ in the inner region where it was expanding freely, up to the termination shock. We consider a SNR that is young enough that it is still propagating inside this un-shocked stellar wind. The structure of a wind bubble can be more complex, especially as there may be multiple wind phases during the lifetime of a star. We refer the reader to the work by \cite{Dwarkadas2005a,Dwarkadas2007b} for the 1D evolution of a SNR inside a more complex wind bubble (without particle acceleration). 
For diffusive shock acceleration we also need a description of the magnetic field in the wind. We adopt a constant magnetization $\sigma = 0.1$ (see \citealt{Chevalier1994b} and \citealt{Lee2014b}), so that~B goes as $r^{-1}$. Upstream of the forward shock, B~varies over time from $44\:\mu$G to $1\:\mu$G. Other parameters are similar to our previous simulations.

\section{Results}
\label{sec:results}

We show here preliminary results for the CC SNR case, with an effective resolution of $256^3$. We reproduce the previous results for the TN SNR case at the same resolution for comparison.

\subsection{Thermodynamics and thermal emission}
\label{sec:results_TH}

\begin{figure}
\center
\includegraphics[width=\figwidth]{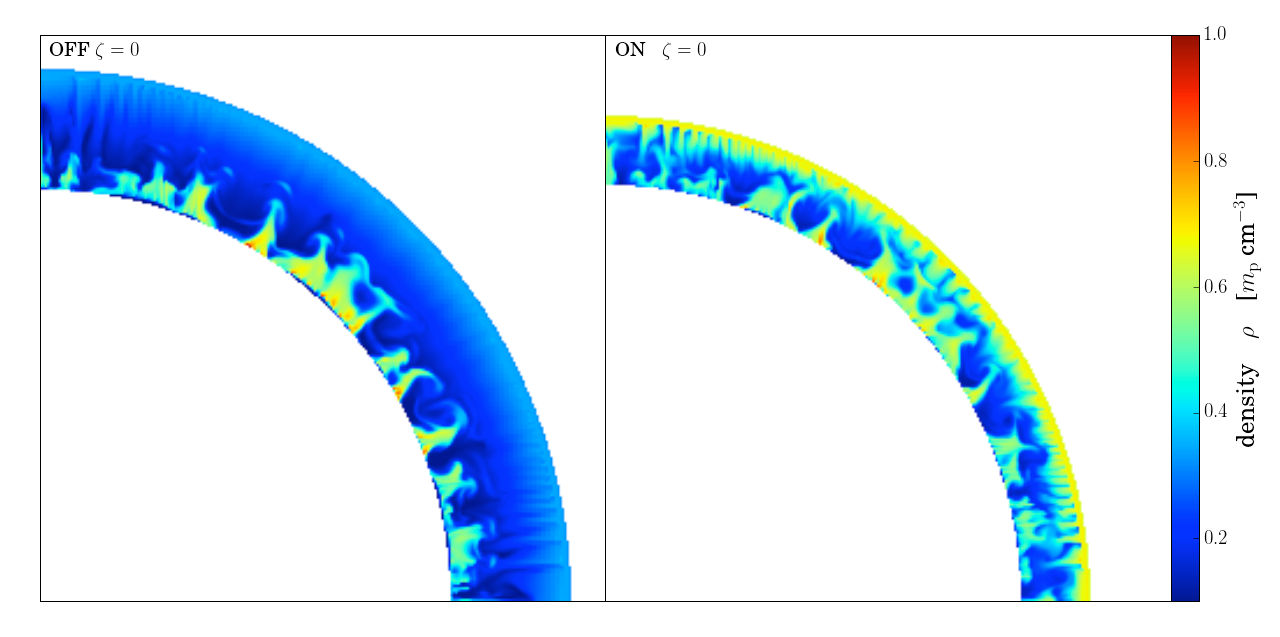}
\includegraphics[width=\figwidth]{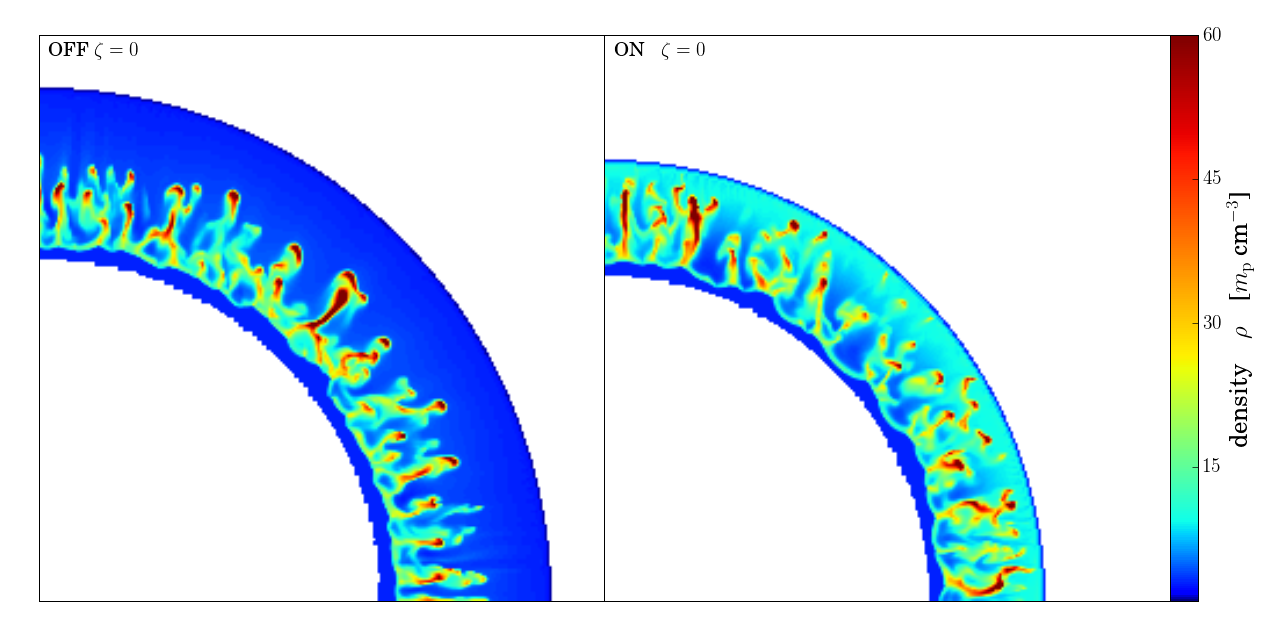} 
\includegraphics[width=\figwidth]{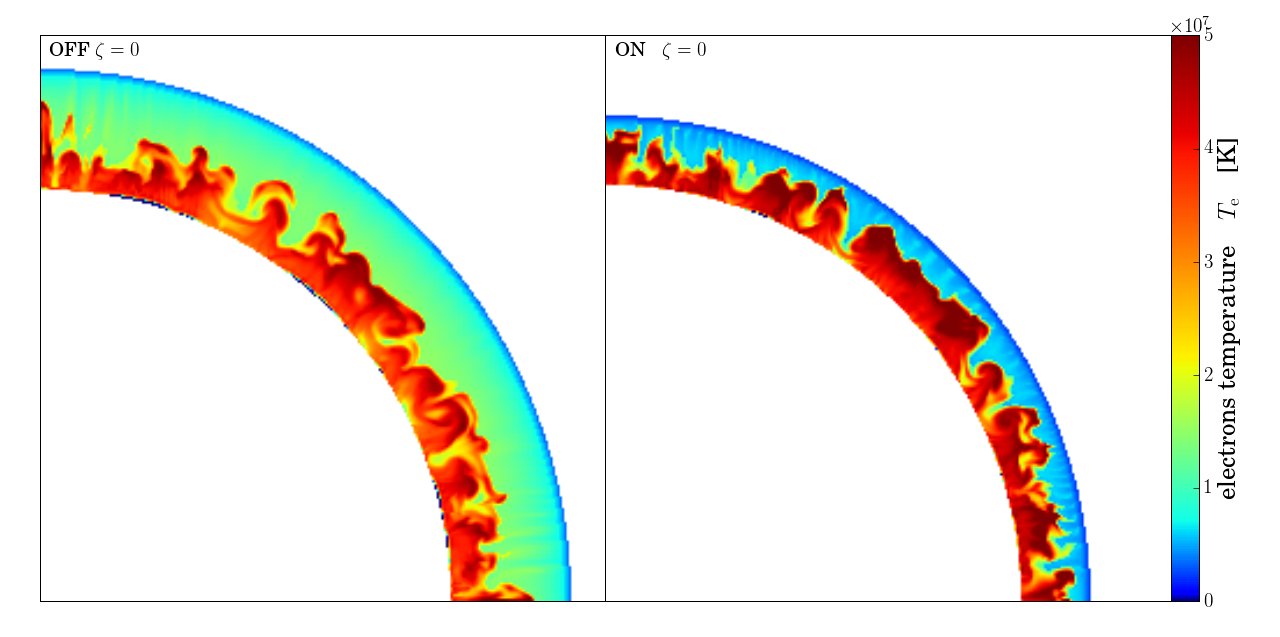}
\includegraphics[width=\figwidth]{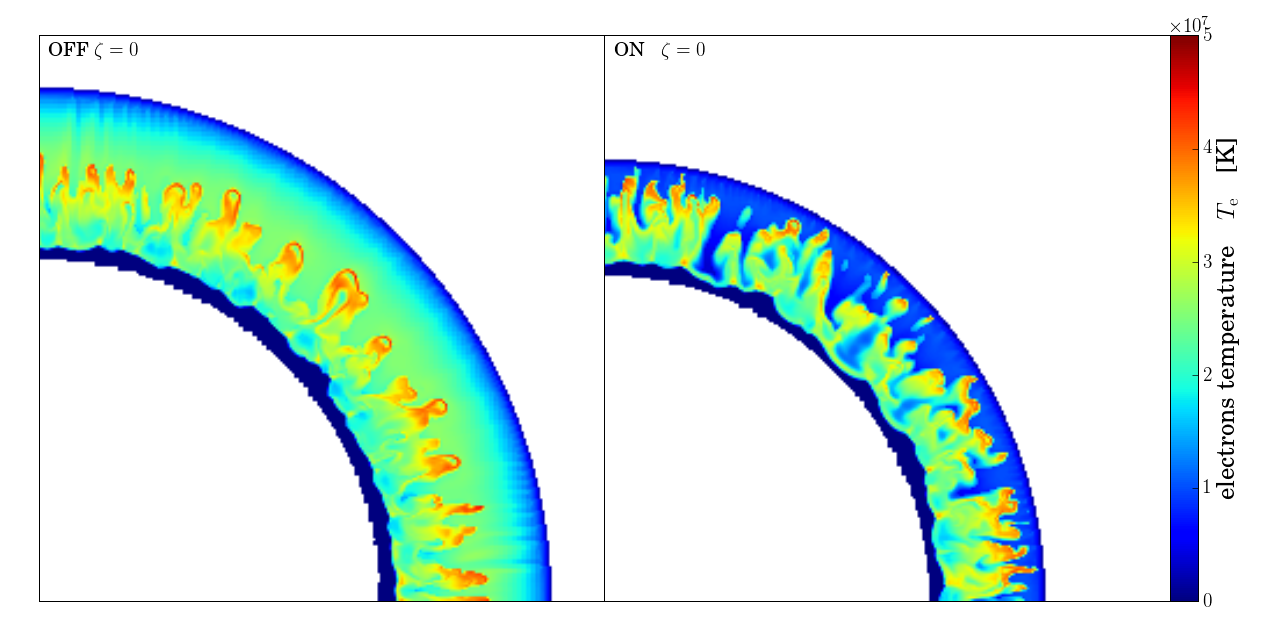}
\caption{Slices in the shocked region. Top: mass density, bottom: electronic temperature. Left: TN SNR, right: CC SNR. Each plot compares the cases without (OFF, on the left) and with (ON, on the right) including the back-reaction of particle pressure.}
\label{fig:maps_hydro}
\end{figure}

\begin{figure}
\center
\includegraphics[width=\figwidth]{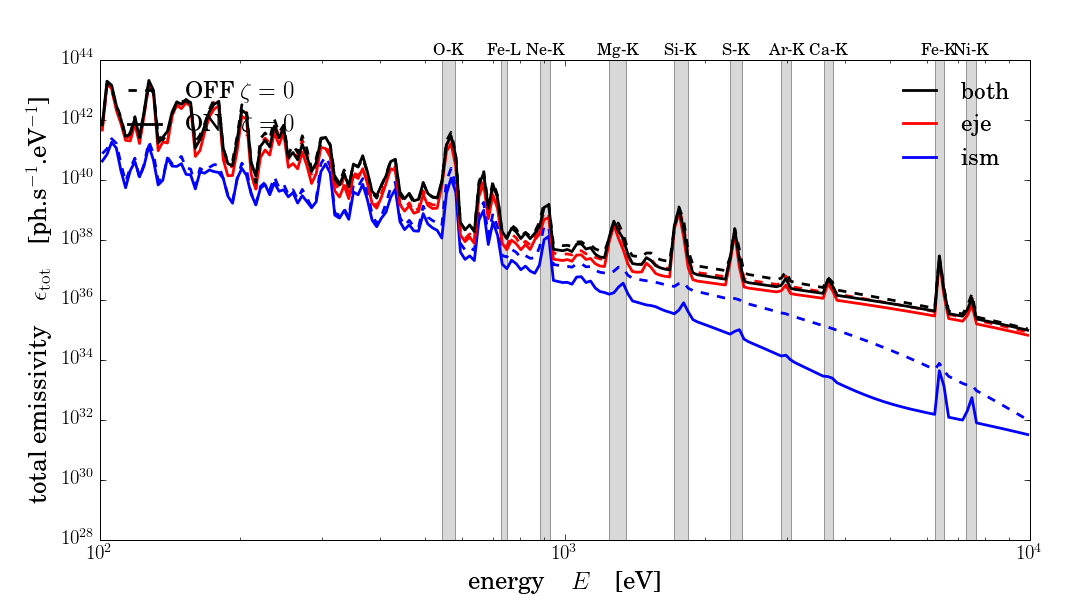}
\includegraphics[width=\figwidth]{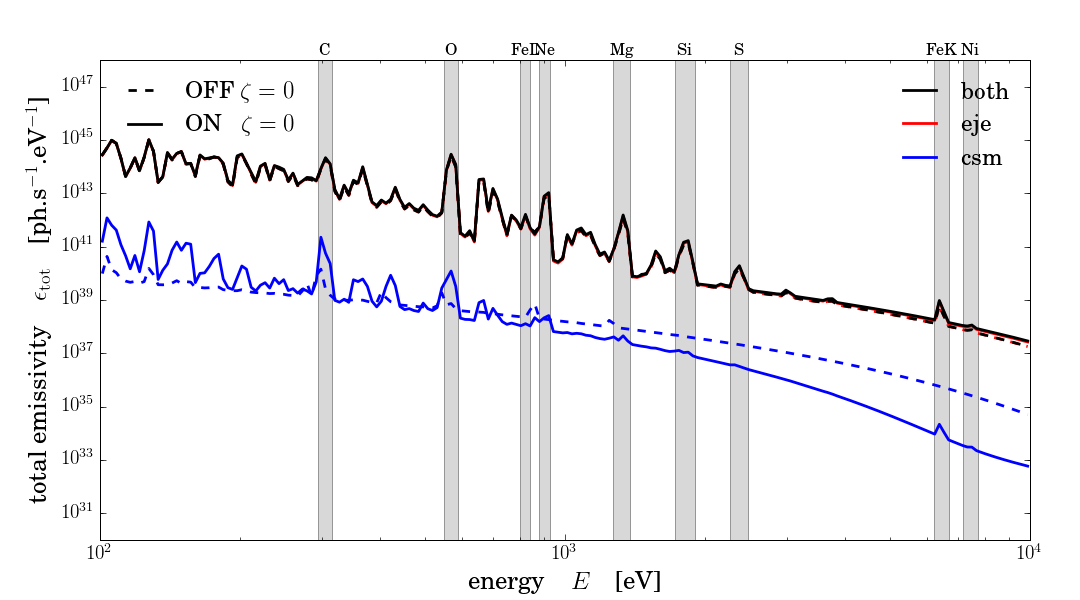}
\caption{Integrated spectra: thermal emission of the plasma. Left: TN SNR, right: CC SNR. Each plot compares the cases without (OFF, dashed) and with (ON, solid) including the back-reaction of particle pressure. The three curves show the emission from the ejecta (in red), from the interstellar/circumstellar medium (in blue), and from both (in black).}
\label{fig:spectra_TH}
\end{figure}

The density in the shocked region is shown on the top of Figure~\ref{fig:maps_hydro}. Compared to the case of a TN SNR (left column), we see that for a CC SNR (right column) the Rayleigh-Taylor fingers are more marked and elongated, and can reach the shock front when the back-reaction is taken into account (``ON'' cases). In the latter case the post-shock protons temperature (not shown here) is greatly reduced as the energy is diverted into particles, and the effect is even more dramatic for a CC SNR. However the electrons are not expected to be heated as fast assuming Coulomb equilibration; we observe a reduction of the post-shock electronic temperature by a factor of about~2. This is illustrated on the bottom of Figure~\ref{fig:maps_hydro}.

The thermal emission of the plasma is computed with a code based on \cite{Mewe1985a}. It depends on the density and temperature, and on the ionization state of the plasma, that has to be computed out of equilibration. For the stellar wind we use solar abundances, for the stellar ejecta we use abundances from a nucleosynthesis model used by \cite{Nozawa2010a} for the modelling of Cas~A. Note that we use a spatially uniform distribution in our simulations, yet we obtain non-uniform thermal emission maps (not shown here) because of the varying plasma conditions. Integrated spectra are presented on Figure~\ref{fig:spectra_TH}, from ejecta-dominated regions, interstellar/circumstellar medium-dominated regions, and both. In both the TN and CC cases, with efficient acceleration the emissivity of the shocked ISM is reduced, all the more so as the photon energy gets higher. But for the CC SNR the total emission is completely dominated by the ejecta, so the effect would be barely detectable.

\subsection{Magnetic field and non-thermal emission}
\label{sec:results_NT}

\begin{figure}
\center
\includegraphics[width=\figwidth]{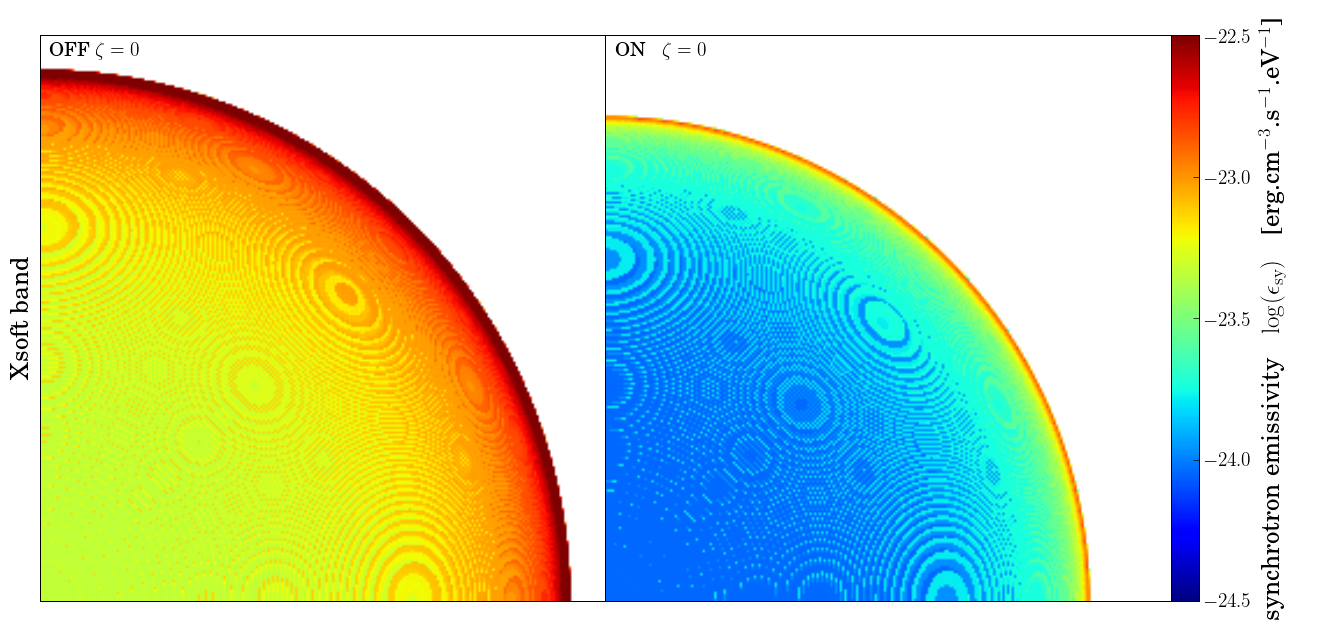}
\includegraphics[width=\figwidth]{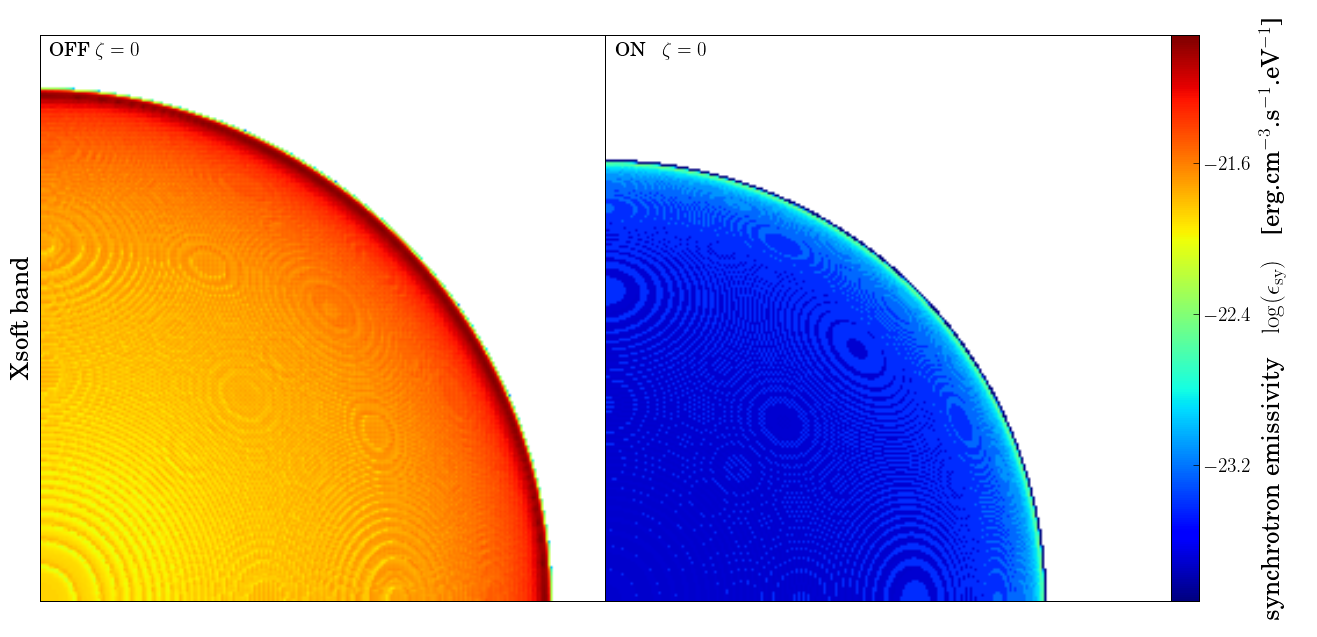}
\includegraphics[width=\figwidth]{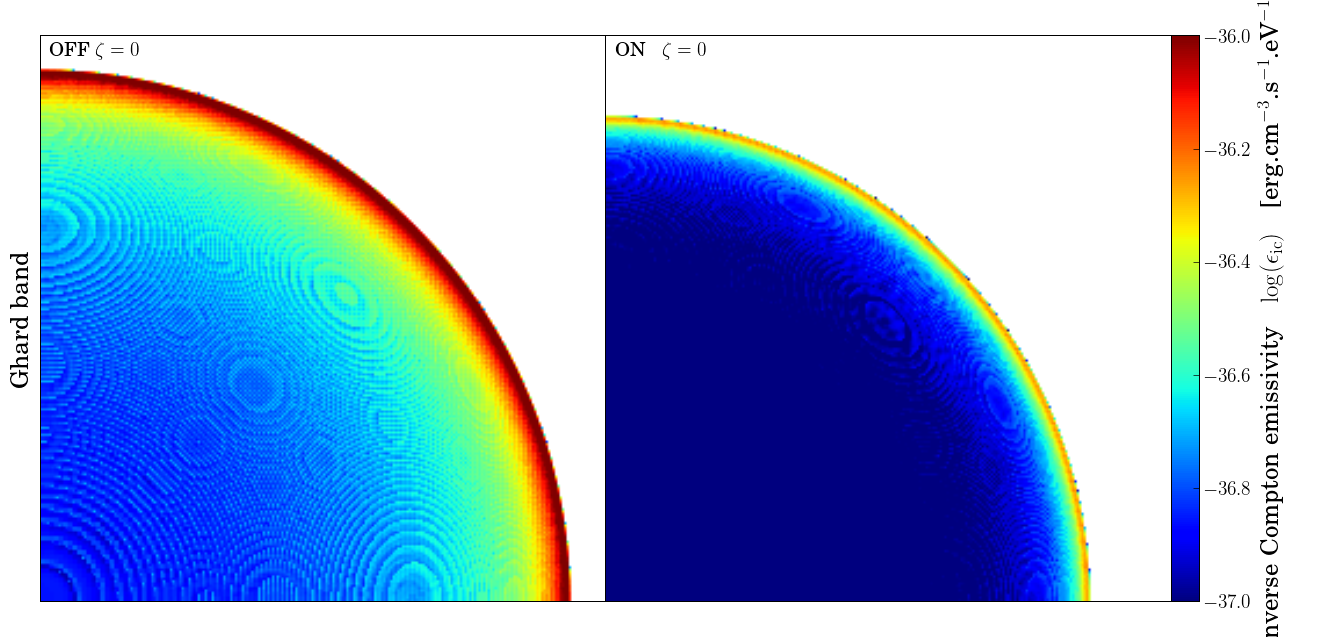}
\includegraphics[width=\figwidth]{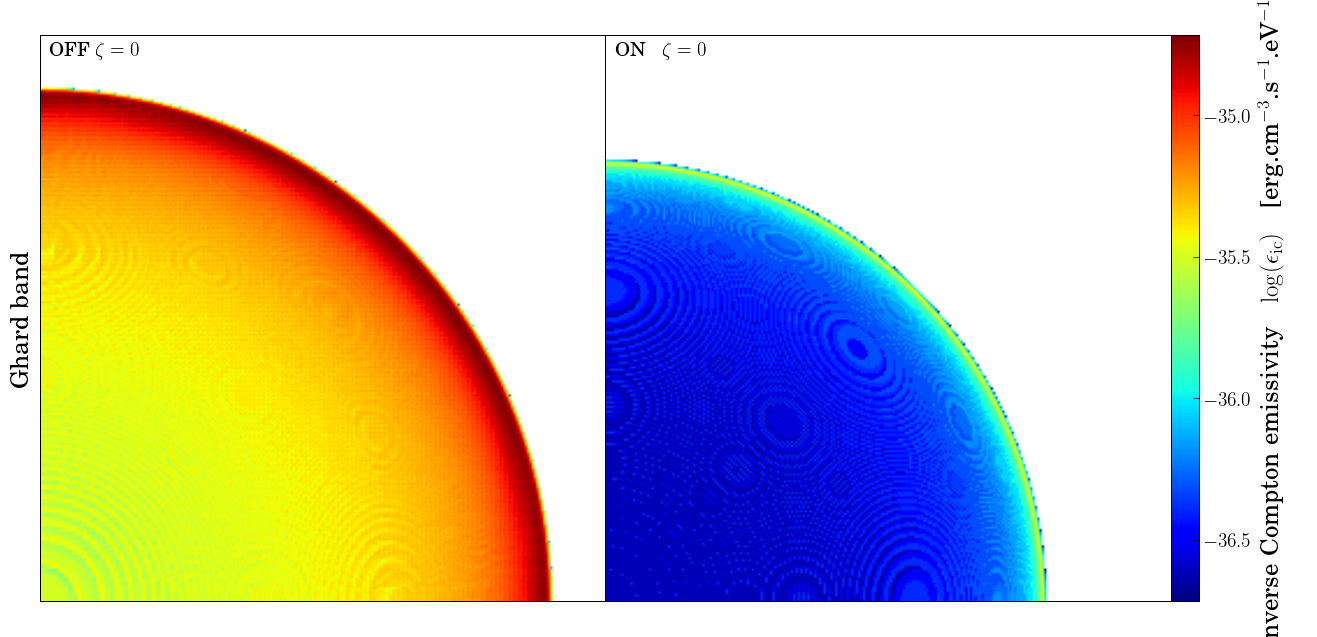}
\includegraphics[width=\figwidth]{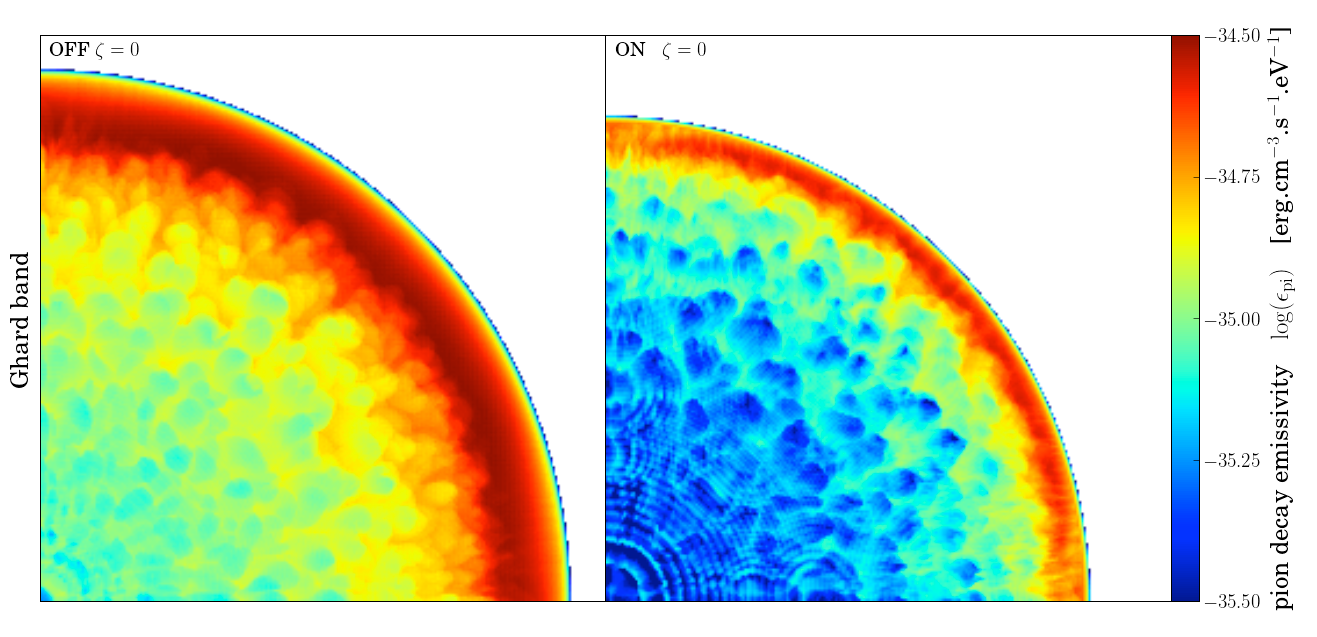}
\includegraphics[width=\figwidth]{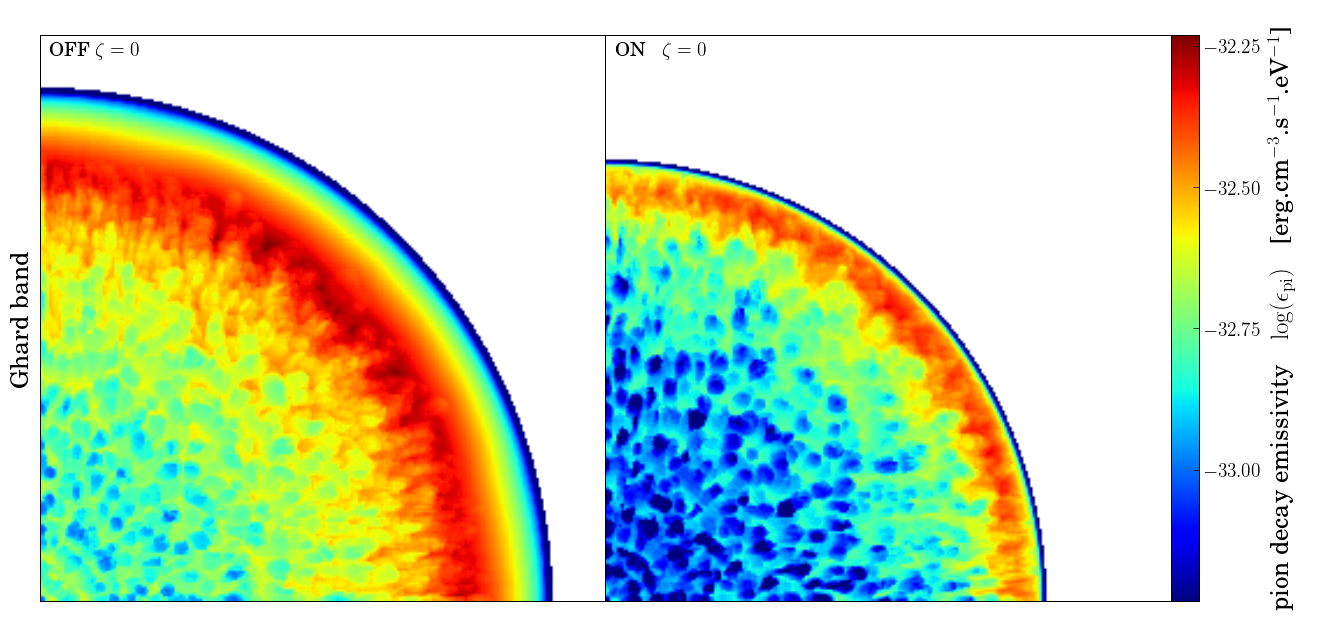}
\caption{Projected maps of the non-thermal emission. Top: X-ray synchrotron, middle: $\gamma$-ray inverse Compton, bottom: $\gamma$-ray pion decay. Left: TN SNR, right: CC SNR. Each plot compares the cases without (OFF) and with (ON) including particle back-reaction.}
\label{fig:maps_NT}
\end{figure}

\begin{figure}
\center
\includegraphics[width=\figwidth]{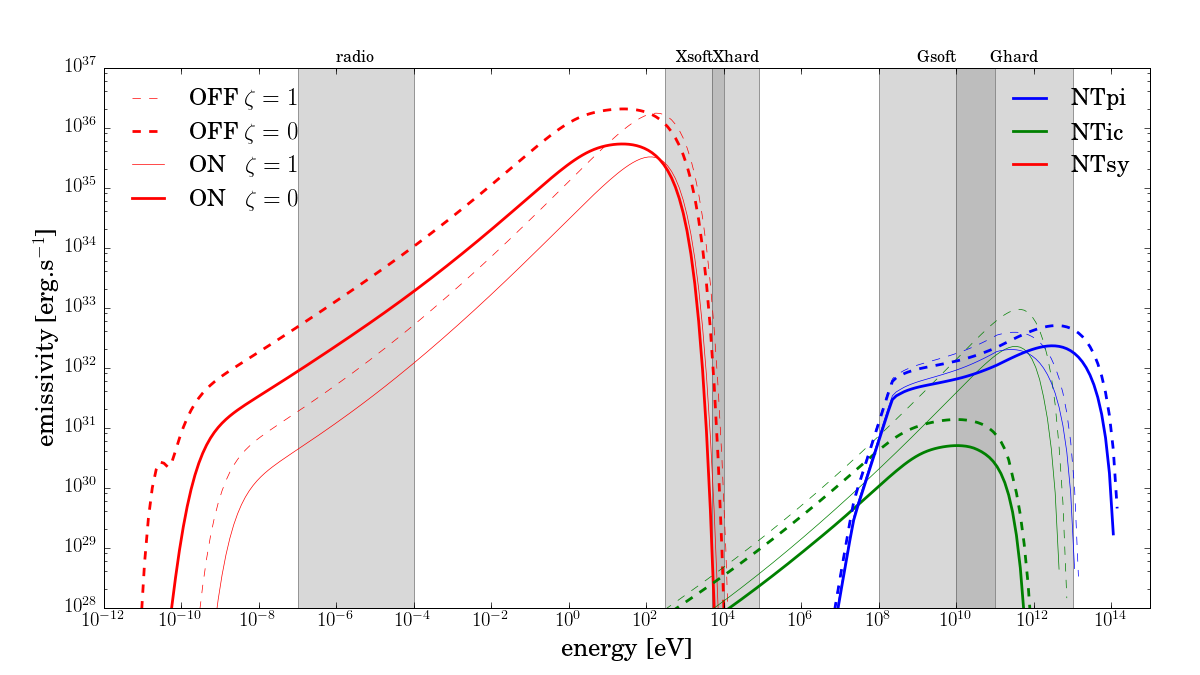}
\includegraphics[width=\figwidth]{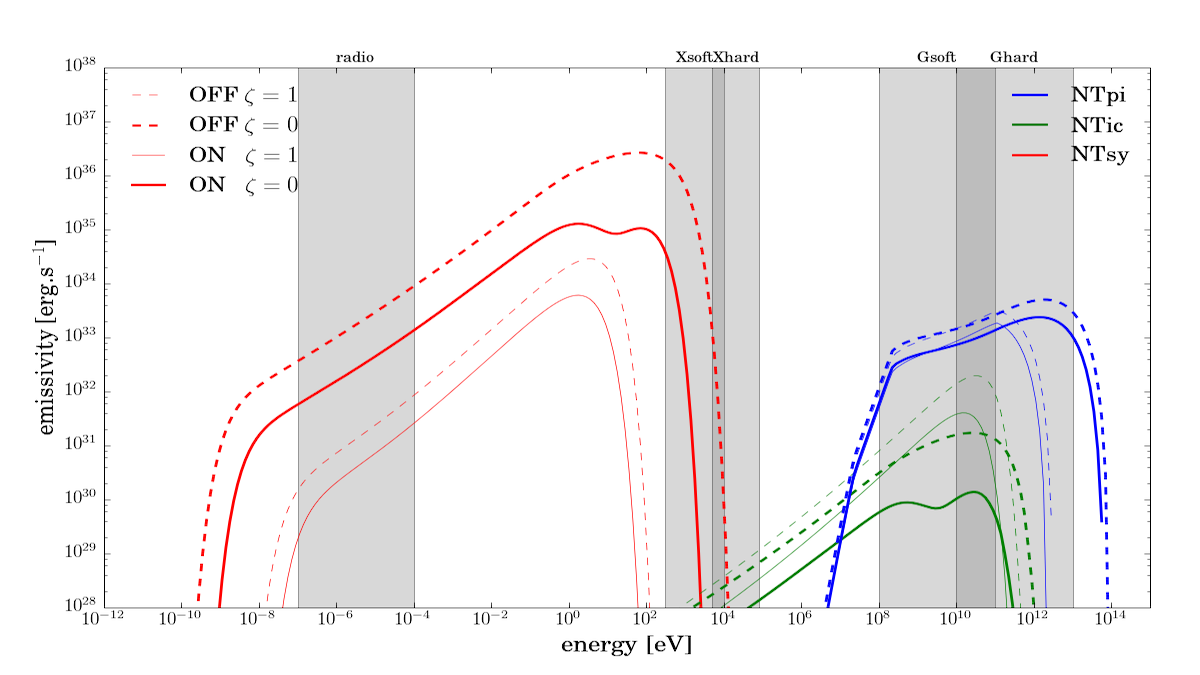}
\caption{Integrated spectra of the non-thermal emission: X-ray synchrotron (red), inverse Compton (green), pion decay (blue). Left: TN SNR, right: CC SNR. Each plot compares the cases without (OFF, dashed lines) and with (ON, solid lines) including the back-reaction of particle pressure, as well as the cases without ($\zeta=1$, thin lines) or with ($\zeta=0$, thick lines) efficient amplification of the magnetic field.}
\label{fig:spectra_NT}
\end{figure}

Particles accelerated at the shock are transported in the downstream region, taking into account losses. Their non-thermal emission is computed with the code AURA, a rewrite of COSMICP \citep{Edmon2011a}. Projected maps are shown on Figure~\ref{fig:maps_NT} in three different bands corresponding to different emission mechanisms, from top to bottom: synchrotron emission from electrons in X-rays, inverse Compton emission from electrons in $\gamma$-rays, and pion decay emission from protons in $\gamma$-rays. The emission from electrons critically depends on the value of the magnetic field downstream of the shock. All the cases presented here were obtained allowing for efficient amplification of the magnetic field in the shock precursor, by a factor of about a hundred according to the model (this corresponds to the parameter $\zeta=0$ in \citealt{Ferrand2014b}). 
In X-rays the synchrotron emission is concentrated in thin rims, as observed. In $\gamma$-rays the inverse Compton leptonic emission is also concentrated in thin rims, although this cannot be observed with current instruments, and no structures are visible inside the remnant. In contrast the hadronic emission peaks behind the forward shock and is more spread out; it also reveals the imprint of the Rayleigh-Taylor fingers in the interior of the SNR. These main effects are qualitatively similar in the TN and CC cases. 

With current $\gamma$-ray instruments the most useful information is in spectral signatures. The broad-band emission from the SNR is shown on Figure~\ref{fig:spectra_NT}. As previously, these plots compare the cases with (``ON'') or without (``OFF'') including the back-reaction of the particle pressure, and they also compare the cases with ($\zeta=0$) or without ($\zeta=1$) efficient amplification of the magnetic field in the shocked region. The most interesting parts of the spectra are in the emission cut-offs. Contrary to the TN case, the X-ray cut-off may not be loss-limited and therefore depend on the value of~B. Similar to the TN case, at TeV energies, magnetic field amplification helps separate hadronic and leptonic emissions by boosting the former and reducing the latter. 

\section{Conclusion}
\label{sec:conc}

As in our previous simulations, the impact of particle acceleration is dependent on the photon energy observed as well as on the magnetic field evolution assumed. In our current model for a CC SNR it appears to be similar for the particles' emission and less visible for the plasma emission. Note that we have not considered here acceleration at the reverse shock. 
We emphasize that the broad-band emission from a SNR is the result of an integration over space and time of the history of the shock strength, particle acceleration, and magnetic field amplification, which makes it difficult to match observations with a one-zone model. 

\small  
\section*{Acknowledgments} 
%
We thank Takaya Nozawa for providing us with nucleosynthesis data.
This work has been supported by NSERC and CFI in Canada.
The numerical simulations were performed on the Nuit computing cluster at the University of Manitoba.

\def\aap{\rm{A\&A}}
\def\aaps{\rm{A\&AS}}
\def\apj{\rm{ApJ}}
\def\apjl{\rm{ApJ}}
\def\mnras{\rm{MNRAS}}
\bibliographystyle{aj}
\small
\bibliography{proceedings}
\end{document}